\documentclass[10pt, letterpaper]{article}
\usepackage[margin=2.5cm]{geometry} 
\usepackage{lmodern}

\usepackage{amssymb}
\usepackage{amsmath}
\usepackage{graphicx}
\usepackage{physics}

\usepackage{color}

\usepackage{nicefrac}
\usepackage{cite}
\usepackage{url}

\title{Revisiting Claims in ``Black Hole Entropy: A Closer Look"}

\author{Pedro Pessoa$^1$, Bruno Arderucio Costa$^2$, Steve Press{\'{e}}$^{3,4,5}$ \\
$^1$Department of Physics, University at Albany - SUNY,   Albany, NY - USA\\
$^2$Institute of Nuclear Sciences, National Autonomous University of Mexico - UNAM, \\ Mexico City - Mexico\\
$^3$Department of Physics, Arizona State University, Tempe, AZ - USA\\
$^4$Center for Biological Physics, Arizona State University, Tempe, AZ - USA\\
$^5$School of Molecular Sciences, Arizona State University, Tempe, AZ - USA }
\date{}

\begin{document}
\maketitle
\abstract{Here we explain how C. Tsallis' reply (Entropy 2021, 23(5), 630) fails to respond to points raised in (Entropy 2020, 22(10), 1110) and introduces further inconsistencies on the origin of black hole entropy.  In his reply, Tsallis argues that the extensivity of thermodynamical entropy with respect to chosen variables needs to be preserved. Consequently the entropy functional is inadequate for black holes. Here we explain the undesirable consequences of this reasoning on black hole thermodynamics.}



\section{Introduction}\label{Introduction}
In a recent paper~\cite{Tsallis20}, Tsallis asserts that the thermodynamic entropy should always be proportional to the volume. Since the Bekenstein--Hawking (BH) entropy is not proportional to the so-called ``volume'' of the black hole, Tsallis states that this deviation is a consequence of the form of entropy functional used in its derivation and suggests the use of a non-additive entropy functional in its derivation.

Henceforth, \cite{Tsallis20} will be referred to as the `initial paper'. 
In \cite{Pessoa20}, henceforth referred to as the `comment', two of us (P.P. and B.A.C.) rebuked his arguments by explaining that
\begin{itemize}
    \item[i] modifying the entropy functional --- defined below in \eqref{KLentropy} --- makes it unsuitable for the purposes of inference. 
    \item[ii] for a system like a black hole, the entropy is not expected to be proportional to any meaningful notion of volume, and
    \item[iii] BH entropy is, apart from a constant, the only formula compatible with the laws of thermodynamics for stationary black holes. Those laws are derived from arguments based on general relativity and quantum field theory alone, with no appeal to a direct calculation from an entropy functional. Therefore, blaming the form of the entropy functional for the \emph{alleged} problem of BH entropy not being proportional to volume is unsound. 
\end{itemize} 

Here we explain how Tsallis fails to rebuke the arguments put forward in \cite{Tsallis21}, henceforth referred to as the `reply'. 

In order to establish notation for a reader that may not have followed the complete discussion and to guarantee a self-contained presentation, some of the arguments presented in the comment will be repeated.

The principle of maximum entropy consists of assigning and updating probability distributions $p(x)$ defined over a system described by $x \in \mathcal{X}$  by \emph{maximizing} the entropy functional given by

\begin{equation} \label{KLentropy}
    S[p|q] = - \int \dd x \ p(x)\ln{\frac{p(x)}{q(x)}} \ ,
\end{equation}
where $q(x)$ is a distribution sometimes termed a ``prior". The form of the entropy functional \eqref{KLentropy} follows from fundamental considerations of what notions one wishes to introduce into $p(x)$ prior to consideration of the data.
As explained in the comment, once \eqref{KLentropy} is supplemented by constraints --- derived from data --- one can evaluate entropy at its maximum value as a function of the constrained values, rather than a functional of probability distributions. As explained by Jaynes \cite{Jaynes65} -- see also \cite{Presse13b,Caticha} -- when constraints are on conserved quantities of a Hamiltonian dynamics, the value of this function is consistent with the laws of thermodynamics and is termed the thermodynamic entropy.

As can be seen in \cite{Vanslette17,Caticha21} --- and under the following two design criteria (DC) {\emph{subdomain independence}} (DC1) implying that local information should have only local effects (unless otherwise later introduced by the data) and {\emph{subsystem independence}} (DC2) implying that a priori independent subsystems should remain independent (again, unless otherwise later introduced by the data) --- the only appropriate form of the entropy is the one monotonic with \eqref{KLentropy}. Naturally, there is broad literature (e.g. \cite{Shore80,Skilling88,LaCour00,Caticha,Presse13,Presse14,Oikonomou17,Oikonomou19,Caticha21}) that reports that extremizing a different functional is unsuitable for the purposes of inference. These issues are direct consequences of violating the DC.
As explained in the comment, violating the DC means that independent subsystems be forced to be considered interacting even if unwarranted by the data. 	  

The simplest example of black hole which exists in the literature --- the Schwarzschild Black Hole (SBH), which has no charge nor angular momentum --- yields the following relationship between its total energy, $M$ and the surface area of its event horizon $A$:

\begin{equation}
    R_s \doteq \sqrt{\frac{A}{4\pi}} = 2M \ ,
    \label{RM}
\end{equation}
where $R_s$ is the Schwarzschild radius.

Above and in the remainder of the present article we use natural units, i.e, the system of units in which the speed of light $c$, Newton's gravitational constant $G$, the reduced Planck constant $\hbar$, and the Boltzmann's constant $k$ are  all equal to $1$. Since the energy of a SBH, namely $M$ in our units, is proportional to the square root of its area \eqref{RM}, it follows --- {without invoking any notion of entropy whatsoever}--- that the energy cannot be proportional to the black hole volume.  

Relating\footnote{This identification is, apart from the numerical factor $1/4$, originally due to Bekenstein~\cite{Bekenstein73} on the grounds of examining the thermodynamics of black hole exteriors.} the following quantity, $S_\text{BH}$, 

\begin{equation}
S_\text{BH}=\frac{A}{4}
\label{BekHawking}
\end{equation}
to a thermodynamic entropy arises from two ingredients: first, a mathematical analogy between the laws of black hole mechanics, which were derived from general relativity~\cite{Bardeen73} and the zeroth, first, and second laws of thermodynamics. Second, the discovery~\cite{Hawking75} that there is a physically meaningful temperature that can be assigned to quantum-mechanical radiation emitted in a spacetime containing a black hole. Moreover, this temperature has the precise form the analogy evinces, which suggests that the entropy in \eqref{BekHawking} is the thermodynamic entropy of a black hole~\cite{Hawking76}. Indeed, one runs into trouble with ordinary thermodynamics on the exterior unless the entropy of a black hole is accounted for~\cite{Bekenstein73}. A striking example thereof is examined by Unruh and Wald~\cite{Unruh82}, based on a system originally conceived by Geroch in an unpublished remark during a conference.

The bottom line is that the formula \eqref{BekHawking} represents an entropy that, when assigned to black holes in general relativity, preserves the laws of thermodynamics\footnote{Here we mean the zeroth, first, and second laws. The third law of thermodynamics has multiple nonequivalent formulations, not all of which are verified in black hole thermodynamics.} to systems containing a black hole.

Note that these macroscopic arguments arriving at a macroscopic formula for black hole thermodynamic entropy do not require maximizing a functional of $p(x)$ --- or, equivalently, statistical mechanics --- as it is unclear what the microscopic degrees of freedom represented by a distribution $p(x)$ within a black hole would be. 

Nor is there any direct calculation in the literature starting from \eqref{KLentropy} leading to the expression \eqref{BekHawking} for a black hole without additional hypotheses on what a microscopic theory of gravity might entail. 

In the reply, Tsallis specifies that his definition of volume of a black hole means that volume $V$ is proportional to $A^{\frac{d}{2}}$ for a black hole in $d+1$ spacetime dimensions. It is worth pointing out, though, that we do not see geometrical meaning in this definition. This is mostly due to the coordinate $r$, which localizes the horizon at $r=R_s$, being timelike for $r<R_s$. For a much more meaningful (time-dependent) notion of the interior volume of a black hole see \cite{Christodoulou15}. {Note that for Tsallis's proposed definition of volume, a new dimension-full parameter must be introduced. Without which, the volume of a $(d-1)$-dimensional hypersurface is proportional to $A^\frac{d}{d-1}$. This exponent of $A$ only coincides with Tsallis's for $d=3$. Violent departures from the well-tested vacuum general relativity are required to make sense of the case $d\neq3$.}

\section{Response}
In this section we will respond point by point the claims made in the reply. 
Although our emphasis is not on terminology, there is a fundamental disconnect between what we and Tsallis refer to as ``thermodynamic entropy". In his reply, Tsallis specifies properties that a quantity ought to obey in order to deserve the name of ``thermodynamic entropy'', among which, a critical one is the Legendre structure. However, Tsallis' understanding of the Legendre structure is unusual, as he does not mean the mere existence of a Legendre transformation between the entropy seen as a function of certain variables, which is an automatic property of concave functions. Instead, Tsallis demands that these variables be extensive, in accordance to the axioms of, for example, the first chapter of Callen's textbook on thermodynamics \cite{Callen}.  We do not see sufficient motivation behind seeking a system consisting of a black hole and perhaps a thermal bath to satisfy this property. Black holes do not posses such a structure, as we shall discuss in Sec.~\ref{Jeopardizing}. Rather, the only relevant thermodynamic property are the very laws of thermodynamics \cite{Hawking76}.

\subsection{BH entropy}\label{Ignoring}

In the reply, Tsallis states:

\begin{quote}
    ``However, it is seemingly undeniable that the sort of perplexity expressed in [49--52], and elsewhere, emerges because, if black holes are thought to be $d=3$ objects, their thermodynamical entropy should be proportional to the cube of the radius, and not to its square, as it happens with Boltzmann–Gibbs-based Bekenstein–Hawking entropy. [\ldots] 
    What is basically argued in [2,3] is that the dominant term of the thermodynamical entropy $S$ of a $d$-dimensional black hole whose event horizon area is $A_H$ is expected to satisfy $\frac{S}{k} \propto\left(\frac{\sqrt{A_H}}{{L_P}}\right)^d.$ Therefore, if $d=2$, we recover the Bekenstein–Hawking entropy, but, if $d\neq 2$, a non-BG entropic functional must be used for thermodynamical purposes."
\end{quote}
As explained in the comment and in Sec. \ref{Introduction}, the BH formula is identified as the thermodynamic entropy of black holes because it leads to the laws of thermodynamics \cite{Bekenstein73, wald1994, Wald01}. Unlike claimed in the reply, the BH formula does not follow from \eqref{KLentropy}. Hence, substituting the entropy \eqref{KLentropy} for another functional will not bear any consequences on the BH formula \eqref{BekHawking}.

\subsection{Jeopardizing classical thermodynamics}\label{Jeopardizing}
Tsallis states in the reply:

\begin{quote}
    ``We read in [1] “\emph{even} if the entropy were proportional to the total energy, it could still fail to be proportional to the “volume” of the black hole.” Such a sentence jeopardizes the Legendre structure of classical thermodynamics, which obviously imposes that \emph{all of its terms scale with size in exactly the same manner}.
    Therefore, the quantities to be legitimately compared are $U,TS,pV,\mu N,H M$, etc. 
    The assumption in [1] about the possibility of the entropy being proportional to the total energy is equivalent to \emph{a priori} assuming that $T$ is intensive, a hypothesis which rather naively disregards that this issue is a very delicate one, given that, in black holes, we definitively deal with long-range interactions."
\end{quote}
It is clear from \eqref{RM} that energy and Tsallis' definition of a black hole's volume are not proportional to each other.

Therefore, if anything jeopardizes Tsallis' understanding of the Legendre structure of classical thermodynamics, it is the existence of the black hole itself. Per \eqref{RM}, the energy and the ``volume'' of a black hole do not scale in the same manner. This seriously raises the questions of which variables Tsallis wishes to build the Legendre structure on and why this is so desirable that he is willing to abdicate either DC1 or DC2 in its favor.

It is also interesting to see that the argument found in the reply is self-contradictory. Tsallis says that the quantities to be legitimately compared are $U, TS, \text{and} \ pV$. In order to give the best possible interpretation of the reply, we assume that $U$ refers to the internal energy (equivalent to $M$ in natural units), $T$ to temperature and $p$ to pressure; even though none of these terms were defined neither on the reply nor in the initial paper. Following Tsallis' understanding, entropy being proportional to energy assumes that temperature $T$ is intensive. Interestingly, following the same logic, entropy being proportional to volume assumes $p/T$ to be intensive. Since no arguments in support of this claim are found in the reply nor on the initial paper, we can see that Tsallis contradicts his own motivations for saying that entropy should be replaced by a non-additive functional. 

Finally, it is useful to point out that a black hole with angular momentum --- referred to as Kerr black hole --- also conflicts with both Tsallis' understanding of the Legendre structure and his proposed solution. The first law of black hole thermodynamics in this case reads

\begin{equation}
    T_{H}\dd S_{BH}=\dd M-\Omega\dd J \ ,
\end{equation}
where $T_H=\frac{1}{4\pi M}\frac{\sqrt{M^2-(J/M)^2}}{M+\sqrt{M^2-(J/M)^2}}$ is the Hawking temperature, $J$ is the angular momentum of the black hole, and $\Omega$ can be thought of as an angular velocity of the horizon. Hence, $\Omega\dd J$ plays a similar role as $-P\dd V$ in ordinary thermodynamics.
For these black holes, the surface area  is~\cite{Wald84} 

\begin{equation}\label{area}
    A=8\pi\left(M^2+\sqrt{M^4-J^2}\right) \ ,
\end{equation}
from which we see that $T_H S_{BH}=\frac{1}{2}\sqrt{M^2-J^2/M^2}$, $M$ and $\Omega J$ cannot all scale in the same manner: if we suppose there is a quantity $S_T$ fitting Tsallis' criterion for entropy, $S_T \propto A^{\frac{d}{2}}$, it follows from \eqref{area} that  $T_H S_T$ still does not scale in $M$ and $\Omega J$ in the same manner. This means that $S_T$ would fail to obey both the first law of thermodynamics \emph{and} Tsallis' understanding of the Legendre structure for a Kerr black hole.

\subsection{Design Criteria}\label{DC}

Tsallis states:
\begin{quote}
    ``Pessoa and Costa apparently base their conviction on the Pressé et al. interpretation of the Shore and Johnson axioms for statistical inference. It happens, however, that they are seemingly unaware that such an interpretation is deeply erroneous."
\end{quote}
Unlike claimed, the arguments in the comment were that DC shouldn't be violated and not directly related to the arguments from Pressé et al. \cite{Presse13}. Tsallis proceeds by presenting his arguments in disagreement to \cite{Presse13}. However, it must be said that contradicting the arguments in \cite{Presse13} is not an argument to say that the DC should be violated.

The only instance in the reply in which Tsallis mentions violating the DC is when writing:

\begin{quote}
    ``It happens, however, that, as it becomes clear within the discussion by Jizba and Korbel in [111], DC1 and DC2 are sufficient but not necessary criteria for the general Shore and Johnson axioms for statistical inference.  For a system with very strong space–time entanglement, such as a black-hole, hypotheses DC1–DC2 are unnecessarily restrictive."
\end{quote}
The argument that the DC are sufficient but not necessary criteria for the general Shore and Johnson axioms misses the point. 
The DC are a mathematical device that allows us to be agnostic as to the coupling between subsystems until such couplings are imposed by data through constraints.

The brief comment on entanglement made by Tsallis quoted above indicates an unfortunate misunderstanding: the DC --- as well as Shore and Johnson axioms --- focus on what properties the entropy functional must obey \emph{before maximization} in order to lead to consistent statistics. The system's scaling properties as well as the correlations following from entanglement refer to the probabilities obtained after maximization \cite{Presse15}.

\subsection{Maximization}
The last sentence in Tsallis' reply reads:
\begin{quote}
    ``a misprint appears above Equation (1) in [1], which reads ``maximization”, but should read ``minimization”."
\end{quote}
This is incorrect. From \eqref{KLentropy}  -- which is the same equation quoted above -- it can be seen trivially that $S$ reduces to Shannon entropy when $q(x)$ is uniform, which is maximized instead of minimized. Moreover, it can be seen from direct calculation that $\fdv[2]{S}{\rho} \leq 0$, hence $S$ is concave and its extreme is a maximum, not a minimum\footnote{It must be said that other works -- e.g. \cite{Nielsen20} -- define Kullback-Leibler divergence as $D_{KL} =  \int \dd x \ p(x)\ln{\frac{p(x)}{q(x)}}$ or, in the notation established here $D_{KL} = -S$. Under this definition, maximizing entropy is equivalent to minimizing KL divergence, $\fdv[2]{D_{KL}}{\rho} \geq 0$ and the extreme for $D_{KL}$ is a minimum. Discussing this terminology is, nevertheless, unnecessary since the comment was self-consistent on that regard.  }.

\section{Conclusions}

Considering the main arguments in the comment -- explicitly written as points (i), (ii) and (iii) in Sec. \ref{Introduction} --- we explained in  Sec. \ref{DC} how the reply evades point (i). In Sec. \ref{Jeopardizing} we explain how the reply fails to appreciate that point (ii) is a clear consequence of general relativity. Finally, in Sec. \ref{Ignoring}, we emphasize how the reply ignores point (iii).

Further comments on disconnection between  Tsallis and us about what qualifies as ``thermodynamic entropy" are in order. 
According to Tsallis, a thermodynamical entropy \emph{has} to imply that $U,TS, \text{and} \ pV$ scale in the same manner. The present article reviews that this property is \emph{not} verified in black hole physics and would require radical modifications in gravitational physics to be rescued.
We, on the other hand, understand that the macroscopic ``thermodynamic entropy" must be the one that enters the laws of thermodynamics. 
Based on that understanding, the entropy obtained after maximizing  \eqref{KLentropy} is the thermodynamic entropy in a Hamiltonian dynamics \cite{Jaynes65}. Following the same logic, the Bekenstein-Hawking entropy is assigned to a black hole without contradicting any core thermodynamic assumptions.


\bibliographystyle{unsrt}
\bibliography{referencias}

\begin{thebibliography}{10}

\bibitem{Tsallis20}
C.~Tsallis.
\newblock Black hole entropy: A closer look.
\newblock {\em Entropy}, 22:17, 2019.

\bibitem{Pessoa20}
P.~Pessoa and B.~Arderucio~Costa.
\newblock {Comment on Tsallis, C. B}lack hole entropy: A closer look.
\newblock {\em Entropy}, 22:1110, 2020.

\bibitem{Tsallis21}
C.~Tsallis.
\newblock Reply to {Pessoa, P.; Arderucio Costa, B.}, {Comment on
  {\textquotedblleft}Tsallis, C. B}lack hole entropy: A closer look. entropy
  2020, 22, 17{\textquotedblright}.
\newblock {\em Entropy}, 23:630, 2021.

\bibitem{Jaynes65}
E.~T. Jaynes.
\newblock {Gibbs vs Boltzmann entropies}.
\newblock {\em American Journal of Physics}, 33:391, 1965.

\bibitem{Presse13b}
S.~Press{\'{e}}, K.~Ghosh, J.~Lee, and K.~A. Dill.
\newblock Principles of maximum entropy and maximum caliber in statistical
  physics.
\newblock {\em Reviews of Modern Physics}, 85:1115, 2013.

\bibitem{Caticha}
A.~Caticha.
\newblock {\em {Entropic Physics: Probability, Entropy, and the Foundations of
  Physics}}.
\newblock 2012.
\newblock Available at:
  \url{https://www.arielcaticha.com/my-book-entropic-physics}.

\bibitem{Vanslette17}
K.~Vanslette.
\newblock Entropic updating of probabilities and density matrices.
\newblock {\em Entropy}, 19:664, 2017.

\bibitem{Caticha21}
A.~Caticha.
\newblock Entropy, information, and the updating of probabilities.
\newblock {\em Entropy}, 23:895, 2021.

\bibitem{Shore80}
J.~Shore and R.~Johnson.
\newblock Axiomatic derivation of the principle of maximum entropy and the
  principle of minimum cross-entropy.
\newblock {\em IEEE Transactions on information theory}, 26:26, 1980.

\bibitem{Skilling88}
J.~Skilling.
\newblock The axioms of maximum entropy.
\newblock In {\em Maximum-Entropy and Bayesian Methods in Science and
  Engineering}, page 173. Springer Netherlands, 1988.

\bibitem{LaCour00}
B.R.~La Cour and W.~C. Schieve.
\newblock Tsallis maximum entropy principle and the law of large numbers.
\newblock {\em Phys. Rev. E}, 62:7494, 2000.

\bibitem{Presse13}
S.~Press{\'{e}}, K.~Ghosh, J.~Lee, and K.~A. Dill.
\newblock Nonadditive entropies yield probability distributions with biases not
  warranted by the data.
\newblock {\em Phys. Rev. Letters}, 111:180604, 2013.

\bibitem{Presse14}
S.~Press{\'{e}}.
\newblock Nonadditive entropy maximization is inconsistent with bayesian
  updating.
\newblock {\em Phys. Rev. E}, 90:052149, 2014.

\bibitem{Oikonomou17}
T.~Oikonomou and G.~Baris Bagci.
\newblock Misusing the entropy maximization in the jungle of generalized
  entropies.
\newblock {\em Physics Letters A}, 381:207, 2017.

\bibitem{Oikonomou19}
T.~Oikonomou and G.~Baris Bagci.
\newblock R{\'{e}}nyi entropy yields artificial biases not in the data and
  incorrect updating due to the finite-size data.
\newblock {\em Phys. Rev. E}, 99:032134, 2019.

\bibitem{Bekenstein73}
J.~D. Bekenstein.
\newblock Black holes and entropy.
\newblock {\em Phys. Rev. D}, 7:2333, 1973.

\bibitem{Bardeen73}
J.~M. Bardeen, B.~Carter, and S.~W. Hawking.
\newblock {The four laws of black hole mechanics}.
\newblock {\em Commun. Math. Phys.}, 31:161, 1973.

\bibitem{Hawking75}
S.~W. Hawking.
\newblock Particle creation by black holes.
\newblock {\em Commun. Math. Phys.}, 43:199, 1975.
\newblock [Erratum: Commun.Math.Phys. 46:206 (1976)].

\bibitem{Hawking76}
S.~W. Hawking.
\newblock Black holes and thermodynamics.
\newblock {\em Phys. Rev. D}, 13:191, 1976.

\bibitem{Unruh82}
W.~G. Unruh and R.~M. Wald.
\newblock Acceleration radiation and the generalized second law of
  thermodynamics.
\newblock {\em Phys. Rev. D}, 25:942, 1982.

\bibitem{Christodoulou15}
M.~Christodoulou and C.~Rovelli.
\newblock How big is a black hole?
\newblock {\em Phys. Rev. D}, 91:064046, 2015.

\bibitem{Callen}
H.~B. Callen.
\newblock {\em Thermodynamics and an introduction to thermostatistics}.
\newblock Wiley and sons, 1985.

\bibitem{wald1994}
R.~M. Wald.
\newblock {\em Quantum field theory in curved spacetime and black hole
  thermodynamics}.
\newblock Chicago Lectures in Physics. University of Chicago Press, 1994.

\bibitem{Wald01}
R.~M. Wald.
\newblock The thermodynamics of black holes.
\newblock {\em Living Reviews in Relativity}, 4:6, 2001.

\bibitem{Wald84}
R.~M. Wald.
\newblock {\em General Relativity}.
\newblock University of Chicago Press, 2010.

\bibitem{Presse15}
S.~Press{\'{e}}, K.~Ghosh, J.~Lee, and K.~A. Dill.
\newblock Reply to {C. Tsallis'} {\textquotedblleft}{C}onceptual inadequacy of
  the {Shore and Johnson} axioms for wide classes of complex
  systems{\textquotedblright}.
\newblock {\em Entropy}, 17:5043, 2015.

\bibitem{Nielsen20}
F.~Nielsen.
\newblock An elementary introduction to information geometry.
\newblock {\em Entropy}, 22:1100, 2020.

\end{thebibliography}


\end{document}